\documentstyle[aps,prb,preprint,epsf]{revtex}
\tightenlines 

\begin{document}
\author{A. G. Rojo$^a$ and C. A. Balseiro$^b$}
\title{Slave fermion theory of confinement in strongly anisotropic systems }
\address{$^a$Department of Physics, The University of\\
Michigan,\\
Ann Arbor, Michigan 48109-1120 \\
}
\address{$^b$Centro At\'omico Bariloche, 8400 Bariloche, Argentina\\
}
\maketitle
\date{The Date }

\begin{abstract}

We present a mean field treatment of a strongly correlated model of 
electrons in a three--dimensional anisotropic system. The mass of the 
bare electrons  is larger in one spatial direction (the $c$--axis direction), 
than in the other two (the $ab$--planes). We use a slave fermion decomposition
of the electronic degrees of freedom and show that there is a transition
from a deconfined to a confined phase in which there is no coherent
band formation along the $c$--axis.

\end{abstract}



\vskip2pc \narrowtext

One of the most controversial, and hard to understand, problems related with
high-$T_{c}$ cuprates is the anomalous charge transport observed
experimentally\cite{THEtheory}. The charge dynamics reflects the anisotropy
in the crystal structure of these compounds, which consists of weakly
coupled planes. In the usual notation, we will refer to ``$c$--axis'' and ``$%
ab$--planes'' as the directions transverse and parallel to the planes
respectively. The in--plane conductivity $\sigma _{ab}$ shows a behavior
characteristic of the metallic state. On the other hand, close to the
insulating state, in the so called underdoped regime, the $c$--axis
conductivity $\sigma _{c}$ is ``incoherent'': the values of $\sigma _{c}$
are below the minimum metallic conductivity\cite{minimum-met}, the
temperature dependence is anomalous, and the frequency dependence does not
show signatures of Drude--like behavior\cite{sigma-freq,sigma2}.

Band structure calculations indicate an anisotropy which, within the
framework of Boltzman transport, imply metallic behavior with an anisotropy $%
\sigma _{c}/\sigma _{ab}$ well above the experimental observation.
Perturbative treatments within the Fermi--liquid theory indicate that the
anisotropy is not renormalized by interactions\cite{shankar}.
Perhaps the main objection to the ``conventional" theories of $c$--axis 
transport\cite{graf,rojo} is the observed value of the anisotropy of the
conductivity.
In the superconducting phase coherence is reestablished 
in all directions\cite{chakravarty}.
This lead
Anderson and others to attribute the anomalies in transport in
the normal state to the effect of
strong electronic correlations, and to conclude that in order to describe
the incoherent $c$--axis conductivity the Fermi--liquid picture should be
abandoned. The starting point used as a paradigm is the one-dimensional
correlated problem where it is rigorously known that the Fermi--liquid
picture fails. Considerable work has been done in weakly coupled chains that
suggest that a state can be formed in which the coherence is confined to the
motion along the chains, the motion transverse to the chains being
incoherent\cite{clarke2}.

A complete theory for the charge dynamics in anisotropic strongly correlated
systems is not yet available. Due to the complexity of the problem, much
work remains to be done in order to develop a fully consistent and
controlled calculation scheme that could account for the phenomenology
indicated by the experiments. In the mean time, the analisys of simple
models is useful as a starting point towards the final answer. Here we
present a mean field treatment of a system of coupled planes that includes
the strong anisotropy, and incorporates the strong correlations responsible
for the non--Fermi liquid behavior. We  show that, within that mean field, a
transition from a deconfined to a confined phase takes place. The parameter
signaling the transition is the gain in kinetic energy due to band formation
in the $c$--axis direction.

We consider the Hubbard model in the limit of infinite on--site repulsion,
described by the following Hamiltonian 
\begin{equation}
H=\sum_{\langle i,j\rangle }t_{i,j}\sum_{\sigma }(1-n_{i,-\sigma
})c_{i,\sigma }^{\dagger }c_{j,\sigma }(1-n_{j,-\sigma }),  \label{H_0}
\end{equation}
where $\langle i,j\rangle $ refers to near neigbors on a cubic lattice where
the anisotropy is incorporated in the values of hopping the matrix elements: 
$t_{i,j}=t_{\Vert }$ for in--plane hoppings and $t_{i,j}=t_{\bot }$ for the
motion along the $c$--axis. The fermion operators $c_{i,\sigma }^{\dagger }$
create an electron at site $i$ only if the site is empty. 

A well known mean field description of Hamiltonian \ref{H_0} is the
slave--boson\cite{anderson,cecatto} approach in which each local
configuration has associated with it a fermionic or bosonic degree of
freedom, such that $c_{i,\sigma }^{\dagger }=a_{i,\sigma }^{\dagger }e_{i}$,
where $a_{i,\sigma }^{\dagger }$ creates a fermion with spin $\sigma $ at
the $i$--th site representing a singly occupied configuration, and $e_{i%
\text{ }}$destroys a boson \ representing the empty state at the same site.
A standard mean field calculation decouples fermions and bosons and relaxes
the exact constraint of one ``particle'' (fermion + boson) per site. The
resulting problem is that of non--interacting bosons and fermions
self--consistently coupled. As a result the ideal bosons condensate in a $%
k=0$ state, the overall effect being a renormalization of the masses of the
fermions. It is important to note that, even for an anisotropic system, the $%
k=0$ bosonic ground state wave function does not know about the anisotropy,
and the mass renormalization is the same in all spatial directions.
Consequently, such an approach preserves the anisotropy and the Fermi liquid
character of the ground state. At least formally, one can conceive
corrections to this state that improve the treatment of the constraint to
avoid multi--occupancy of the particles at the same site. There are,
however, other alternative treatments that--still within the mean field
level--take into account the hard core constraint for the bosons exactly. In
the present work we present a mean field along this line. In what follows we
show that at an alternative description in terms of slave--fermions \ for
the infinite--$U$ case breaks the Fermi liquid description and produces a
confined coherent sate\cite{clarke} in the $ab$ planes.

We introduce a description in which the original projected fermions are represented by
three fermions: 
\begin{equation}
\overline{c}_{i,\sigma }\equiv c_{i,\sigma }(1-n_{i,-\sigma})=a_{i,\sigma }f_{i,\uparrow }^{\dagger }f_{i,\downarrow }
\label{slave}
\end{equation}

The above representation respects the anticommutation relation between the
projected operators $\overline{c}_{i,\sigma }$ and $\overline{c}_{i,\sigma }^{\dagger }$ provided one stays
within the physical Hilbert space.
A related fermion linearization was presented in Ref. \cite{zanardi}.

The product $f_{i,\uparrow }^{\dagger }f_{i,\downarrow }$ is a spin flip
operator corresponding to a pseudo--spin degree of freedom not related to $%
\sigma $. When this fictitious spin is $\downarrow $ in site $i$ this means
that the site is occupied, and the site is empty if the spin is $\uparrow $:
there are as many $f_{\downarrow }$'s as there are electrons and as many $%
f_{\uparrow }$'s as there are holes. The $f$ fermions therefore satisfy

\begin{equation}
\langle f_{i,\uparrow }^{\dagger }f_{i,\uparrow }\rangle +\langle
f_{i,\downarrow }^{\dagger }f_{i,\downarrow }\rangle =1,\text{ }\sum_{\sigma
}\langle a_{i,\sigma }^{\dagger }a_{i,\sigma }\rangle +\langle
f_{i,\downarrow }^{\dagger }f_{i,\downarrow }\rangle =1,  \label{consf}
\end{equation}
and, in turn 
\begin{equation}
\sum_{\sigma }\langle a_{i,\sigma }^{\dagger }a_{i,\sigma }\rangle =1-\delta
,  \label{consa}
\end{equation}
with $\delta $ representing the fractional deviation in occupation number
with respect to the half--filling case of one electron per site.

At the mean field level the ground state wave function consists of a direct
product of three Fermi seas, one per each of the fermion degrees of freedom.
The total energy in this approximation is given by 
\begin{equation}
E_0 = -\sum_{\langle i,j\rangle}t_{i,j} A_{i,j} \chi_{i,j}^2,
\end{equation}
with 
\begin{equation}
A_{i,j} = \sum_{\sigma} \langle a^{\dagger}_{i,\sigma}a_{j,\sigma}\rangle,
\end{equation}
\begin{equation}
\chi_{i,j}=\langle f^{\dagger}_{i,\uparrow}f_{j,\uparrow} \rangle= \langle
f^{\dagger}_{i,\downarrow}f_{j,\downarrow} \rangle,
\end{equation}
where the last equality holds because we are dealing with a bipartite
lattice with particle--hole symmetry. The three species of fermions are free
with their hopping amplitudes renormalized by the factors $A_{i,j}$, and $%
\chi_{i,j}$. These factors are responsible for renormalizing the anisotropy
and can be better visualized  in the mean field Hamiltonian 
\begin{equation}
H_{{\rm MF}} =-\sum_{\langle i,j\rangle}t_{i,j}\sum
_{\sigma}\left[\chi_{i,j}^2 a^{\dagger}_{i,\sigma}a_{j,\sigma} + A_{i,j}
\chi_{i,j}f^{\dagger}_{i,\sigma}f_{j,\sigma} \right] + C,
\end{equation}
with $C$ a constant.

Note that, for small deviations from half filling, the $f_{i,\uparrow}$ ($%
f_{i,\downarrow}$) fermions are moving close to the bottom (top) of their
band, whereas the $a$ fermions are close to the center of the band. This
makes their respective Fermi surfaces different.

Our mean field can be understood in two steps. First the $a$ fermions are
decoupled from the $f$ fermions. At that level, the $a$ fermions are free,
but the $f_{\uparrow }$ fermions and the $f_{\downarrow }$ fermions are
strongly correlated. The dynamics of the system of $f$ fermions at this
level is identical to that of an $xy$--model, and can be mapped onto a
hard--core boson problem. In the second step the $f$ fermions of different
spin are decoupled and treated as free fermions (with a self consistent
constraint on the dynamics)\cite{affleck}. Note that, at the level of step
one, the above mentioned system of hard--core bosons will in principle have
an anisotropy in the expectation values of the kinetic energy terms that
will depend on direction. This is due to the quantum fluctuations introduced
by the hard--core constraint.

At half filling ($\delta =0$), the kinetic energy of the $f$ fermions is
zero, the renormalization factor $\chi _{ij}=0$, giving the localized limit
of the $a$ fermions which we identify as the Mott insulating state. On the
other hand, far from half filling, for $\delta \sim 1$, the density of $a$
--fermions is so low that they should behave as non--interacting, but our
mean field fails to recover this limit. Decoupling the $f$--fermions from
the $a$--fermions is not a good approximation in the limit of high doping $\
\delta $ because the probability of finding an $f_{\downarrow }$-fermion at
a site occupied by an $a$ fermion is very low [$\sim (1-\delta )^{2}$],
while the exact dynamics requires this probability to be one. Therefore our
results will be valid close to half filling, or $\delta \sim 0$.

Due to translational invariance, the ground state energy will be a function
of the four quantities $A_{\|}$, $A_{\bot}$, $\chi_{\|}$ and $\chi_{\bot}$: 
\begin{equation}
E_0= -4t_{\|} A_{\|} \chi_{\|}^2 -2t_{\bot}A_{\bot} \chi_{\bot}^2.
\end{equation}

The one particle energies of the $f$ and $a$ fermions are respectively 
\begin{equation}
E_{{\bf k}}^f =t_{\|} A_{\|} \chi_{\|}\varepsilon_{k_{\|}}-t_{\bot}A_{\bot}
\chi_{\bot} 2\cos k_z,
\end{equation}
\begin{equation}
E_{{\bf k}}^a =t_{\|} \chi_{\|}^2\varepsilon_{k_{\|}} -t_{\bot}
\chi_{\bot}^2 2\cos k_z,
\end{equation}
with 
\begin{equation}
\varepsilon_{k_{\|}}= -2(\cos k_k +\cos k_y).
\end{equation}

Effective chemical potentials $\lambda $ and $\mu$ have to be determined for
each of the two types of fermions through the equations 
\begin{equation}
{\frac{1}{N}} \sum _{{\bf k}} f(E_{{\bf k}}^f-\lambda) =\delta,
\;\;\;\;\;\;\;\; {\frac{1}{N}} \sum _{{\bf k}} f(E_{{\bf k}}^a-\mu) ={\frac{%
1-\delta}{2}}.
\end{equation}

We approximate the reduced density of states corresponding to the motion
within the plane by a constant: $\sum_{k_{\Vert }}\delta (\varepsilon
-\varepsilon _{k_{\Vert }})=\Theta (4-|\varepsilon |)/4$, and find that the
mean field equations can be written in terms of the parameters $\alpha $ and 
$\beta $ defined as 
\begin{equation}
\alpha ={\frac{t_{\bot }}{t_{\Vert }}}{\frac{A_{\bot }\chi _{\bot }}{%
A_{\Vert }\chi _{\Vert }}},\;\;\;\;\;\;\;\;\;\beta ={\frac{t_{\bot }}{%
t_{\Vert }}}\left( {\frac{\chi _{\bot }}{\chi _{\Vert }}}\right) ^{2}.
\label{equ0}
\end{equation}

After straightforward integrations, and using the fact that close to half
filling the Fermi surface of the $a$ fermions is open, the mean field
equations are 
\begin{equation}
A_{\Vert }={\frac{1}{2}}(1-\delta ^{2})-\left( {\frac{\beta }{2}}\right)
^{2}\;,\;\;\;\;\;A_{\bot }={\frac{\beta }{4}},  \label{equ1}
\end{equation}
\begin{equation}
\chi _{\Vert }={\frac{1}{8\pi }}\left\{ \left[ 1-\left( {\frac{\widetilde{%
\lambda }}{4}}\right) ^{2}\right] 2k_{0}-{\frac{\alpha ^{2}}{4}}\left( k_{0}+%
{\frac{1}{2}}\sin 2k_{0}\right) -{\frac{\widetilde{\lambda }\alpha }{2}}\sin
k_{0}\right\} ,
\end{equation}
\begin{equation}
\chi _{\bot }={\frac{1}{4\pi }}\left\{ \left( 1-{\frac{\widetilde{\lambda }}{%
4}}\right) 2k_{0}\sin k_{0}+{\frac{\alpha }{2}}\left( k_{0}+{\frac{1}{2}}%
\sin 2k_{0}\right) \right\} ,  \label{equn1}
\end{equation}
with $\widetilde{\lambda }=\lambda /(t_{\Vert }A_{\Vert }\chi _{\Vert })$
determined from the equation 
\begin{equation}
\delta ={\frac{1}{8\pi }}\left\{ (\widetilde{\lambda }+4)k_{0}+2\alpha \sin
k_{0}\right\} ,  \label{equn}
\end{equation}
and $k_{0}=\cos ^{-1}[-(\widetilde{\lambda }+4)/2\alpha ]$ for $|(\widetilde{%
\lambda }+4)/2\alpha |<1$ and $\pi $ otherwise. Note that $\alpha $ plays
the role of an effective anisotropy of the $f$--fermions. For a given $%
\delta $, if we fix $\alpha $, the renormalization factors $\chi $ and $A$
are determined by the Equations (\ref{equ1}) through (\ref{equn}). This
means that $\alpha $ plays the role of a variational parameter with respect
to which we have to minimize the energy $E_{0}$. As an example, in Figure 
\ref{figura2} we show some curves of $E_{0}$ vs. $\alpha $ for different
values of doping using as a parameter the bare anisotropy $t_{\bot
}/t_{\Vert }$. 

The curves indicate that for fixed $t_{\bot }/t_{\Vert }$ there is a
discontinuous jump in the position of the minimum of $E_{0}$ as $\delta $ is
varied. The curve shown in Figure \ref{figura2} for $\delta =0.002$
corresponds to the confined phase for which $\alpha =0$, and the
renormalization factor $\chi _{\bot }=0$. The curve for $\delta =0.0018$ has
its minumum at finite $\alpha $ and hence corresponds to a three dimensional
metal with a renormalized anisotropy.

A phase diagram that result from our calculation is shown in Figure \ref
{figura3} 

\bigskip

A very important point is to establish that the particle motion does not
correspond to a Fermi liquid. We show this by computing the form of the
occupation number of the original fermions in the confined phase within our
mean--field squeme: 
\[
n_{k,\sigma }\equiv \langle c_{k,\sigma }^{\dagger }c_{k,\sigma }\rangle =%
\frac{n}{2}+\frac{1}{N}\sum_{i\neq j}e^{ik(R_{i}-R_{j})}\langle c_{i,\sigma
}^{\dagger }c_{j,\sigma }\rangle 
\]
with $n$ the particle density. The term $\langle c_{i,\sigma }^{\dagger
}c_{j,\sigma }\rangle $ is evaluated using the representation of Equation 
\ref{slave}. In mean field the result is a convolution of the occupation
numbers of the three fermions ($f$'s and $a$). Using the constraints of
Equations \ref{consf} and \ref{consa} one obtains 
\[
n_{k,\sigma }=\frac{1-\delta }{2}\left[ 1-\delta (1-\delta )\right] +\frac{1%
}{N^{2}}\sum_{pq}n_{p,\sigma }^{(a)}n_{q,\downarrow }^{(f)}n_{p+q-k,\uparrow
}^{(f)}. 
\]

The occupation numbers above correspond to three Fermi surfaces. For small $%
\delta $ the Fermi surfaces corresponding to the $f$ fermions are are two
circles centered respectively at ${\bf k=(}0,0)$ and ${\bf k=(}\pi ,\pi )$.
On the other hand, the Fermi surface of the $a$ fermions are close to a
diamond. The result of the convolution above (See Figure 4) is that $%
n_{k,\sigma }$ does not have a discontinuity, implying an non--Fermi liquid
state.

A few points related to the calculation deserve a comment:

{\it i) }Due to the approximation made in the density of states we cannot
recover the isotropic case. The approximation used is aimed at describing
anisotropic systems.

{\it ii) }In our calculation the confined regime is identified by the
vanishing of the expectation value of the interplane hopping indicating that
there is not band formation along this direction. We interpret this result
as an indication of incoherence, even though one expects some
interplane--coupling to remain in the exact incoherent regime. The picture
is analog to the slave boson description of the Mott insulator. There, the
insulating state is characterized by a vanishing of the inter--site hopping,
while we know that in the exact ground state this magnitude is small but
finite.

In summary, we have presented a mean field calculation and derived a
phase--diagram of a strongly interacting anisotropic system. We have shown
that, as the anisotropy increases, for small deviations from half filling a
transition\ takes place from a deconfined phase to a confined phase in which
the motion in the $c$--axis direction is completely incoherent while the
motion in the $ab$--direction corresponds to a coherent, non--Fermi liquid
state.

\begin{figure}
\epsffile{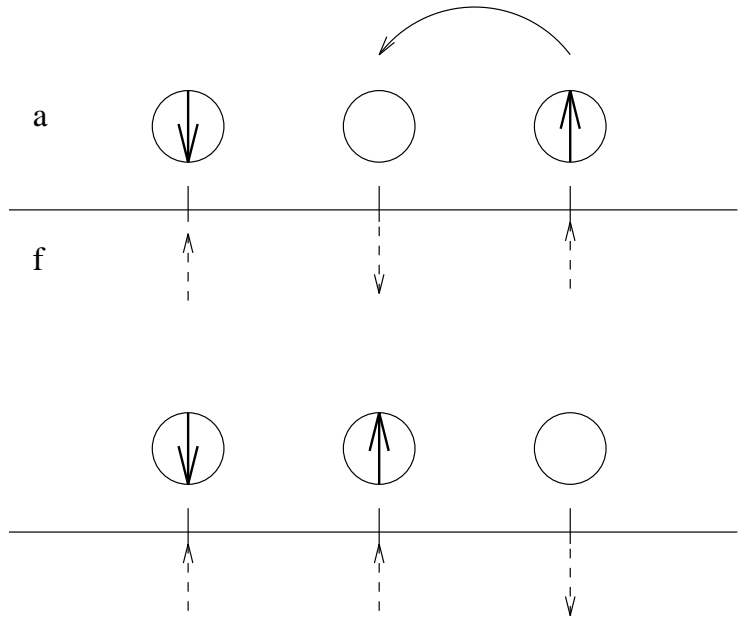}
\caption{
Schematic rendition of the slave fermion decomposition. When an
 $a$ 
   fermion hops from site $i+1$ to site $i$, there is a spin--flip of $f$
   fermions represented by the dashed arrows. The upper (lower) part of the
	 figure represents the configuration before (after) the hopping process.}

\end{figure}

\begin{figure}
\epsffile{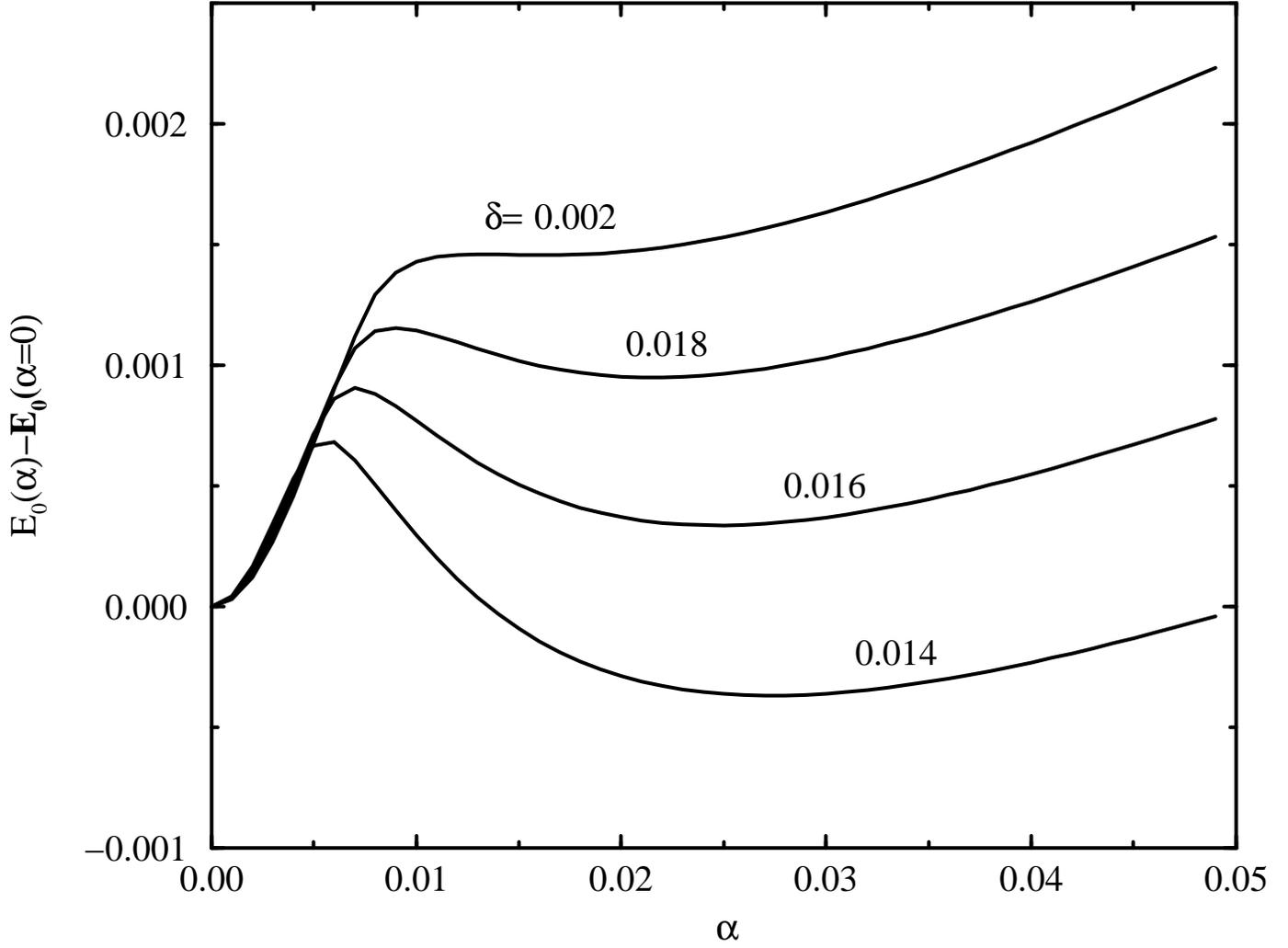}
\caption{ Ground state energy vs. the variational parameter $\alpha $for an
anisotropy of $t_{\bot }/t_{\Vert }=0.3$ and the indicated values of doping.}
\label{figura2}
\end{figure}

\begin{figure}
\epsffile{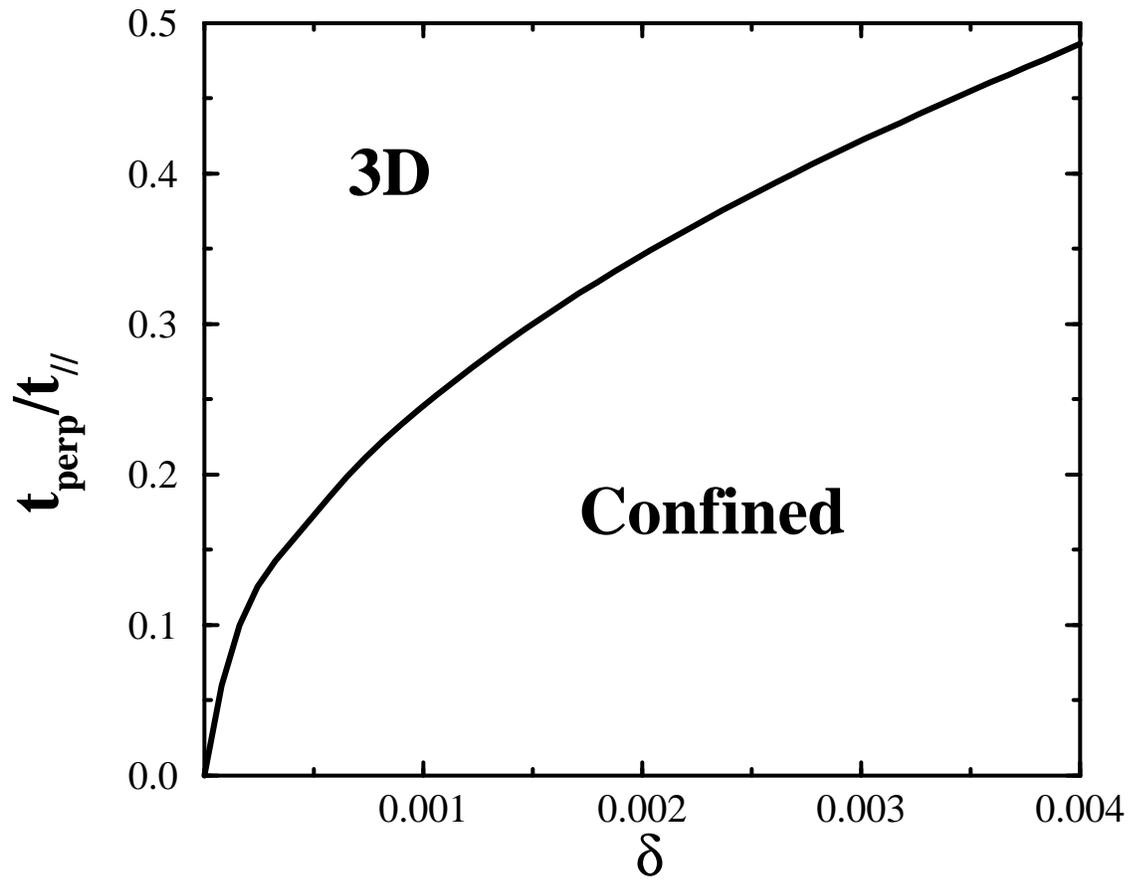}
\caption{ Phase diagram valid in the low doping regime indicating the
boundary between a confined phase and a three--dimensional anisotropic
phase. }
\label{figura3}
\end{figure}
\newpage
\begin{figure}
\epsffile{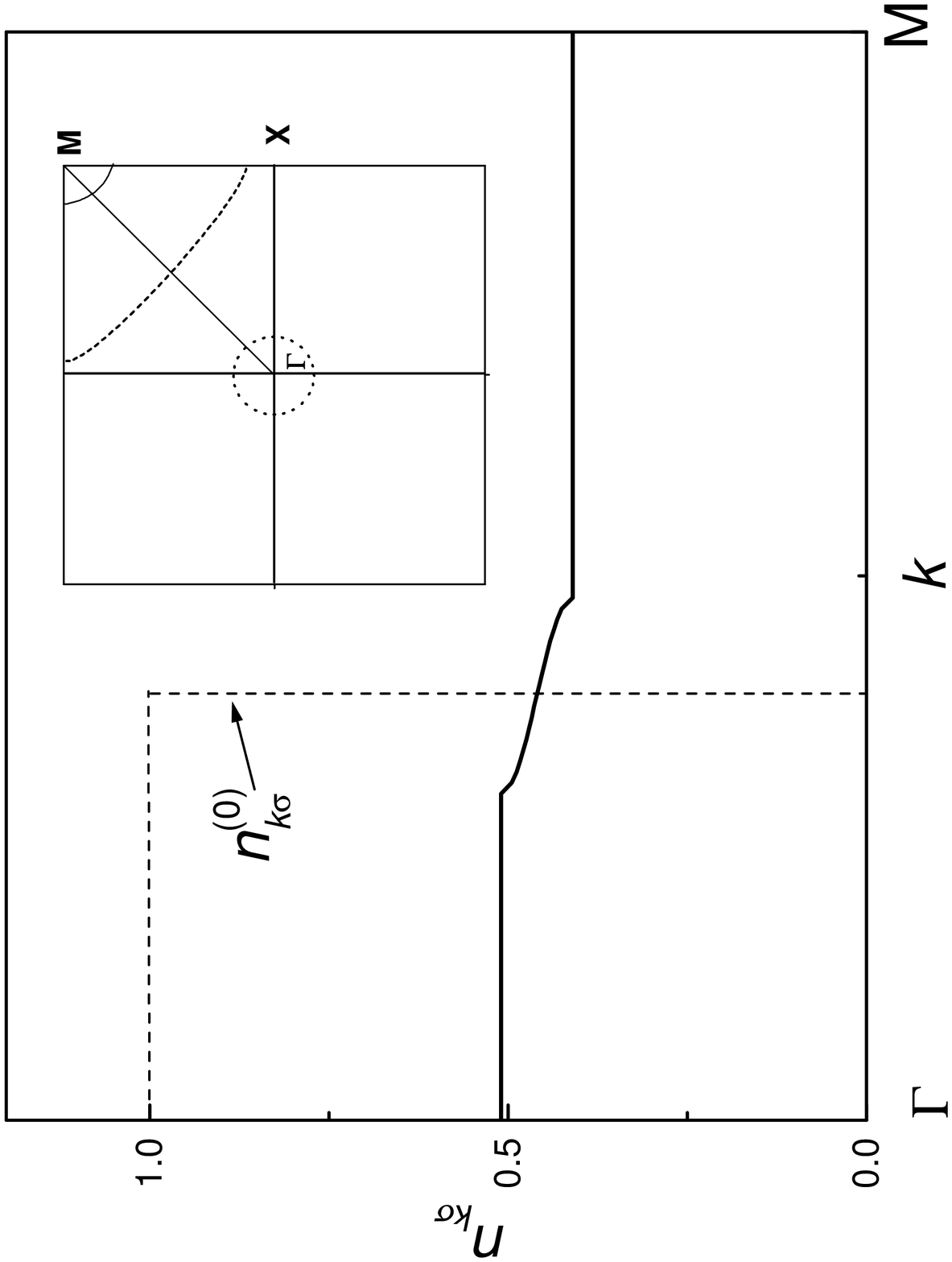}
\caption{Fermion occupation number vs. wave vector in the direction marked
in the inset. The dashed line shows the bare, non--interacting,
occupation number. 
Note that there is no discontinuity in  $n_{k,\sigma}$, as expected in a
non--Fermi--liquid state.
The inset also shows the Fermi surfaces--in the first quadrant
only--of the $f$ fermions (small circle shown in short dashed and 
arc in  continuous line
 close to the $M$ point) and of the $a$ fermions (dashed line).}
\end{figure}
\end{document}